\documentclass[preprintnumbers,showkeys,showpacs,byrevtex]{revtex4} 
\usepackage{amsmath,amsfonts,amssymb,amscd,amsxtra,amsthm}
\usepackage{graphicx}
\usepackage{bm}
\begin{document}   
\preprint{CYCU-HEP-09-09}
\title{An effective thermodynamic potential from the instanton\\
with Polyakov-loop contributions}      
\author{Seung-il Nam}
\email[E-mail: ]{sinam@cycu.edu.tw}
\affiliation{Department of Physics, Chung-Yuan Christian University, Chung-Li 32023, Taiwan} 
\date{\today}
\begin{abstract}  
We derive an effective thermodynamic potential ($\Omega_\mathrm{eff}$) at finite temperature ($T\ne0$) and zero quark-chemical potential ($\mu_{\mathrm{R}}=0$), using the singular-gauge instanton solution and Matsubara formula for $N_{c}=3$ and $N_{f}=2$ in the chiral limit. The momentum-dependent constituent-quark mass is also obtained as a function of $T$, employing the Harrington-Shepard caloron solution in the large-$N_c$ limit. In addition, we take into account the imaginary quark-chemical potential $\mu_{\mathrm{I}}\equiv A_4$, translated as the traced Polayakov-loop ($\Phi$) as an order parameter for the $\mathbb{Z}(N_{c})$ symmsetry, characterizing the confinement (intact) and deconfinement (spontaneously broken) phases. As a result, we observe the crossover of the chiral ($\chi$) order parameter $\sigma^{2}$ and $\Phi$. It also turns out that the critical temperature for the deconfinment phase transition, $T^{\mathbb{Z}}_c$ is lowered by about $(5-10)\%$ in comparison to the case with a constant constituent-quark mass. This behavior can be understood by considerable effects from the partial chiral restoration and nontrivial QCD vacuum on $\Phi$. Numerical calculations show that the crossover transitions occur at $(T^{\chi}_c,T^{\mathbb{Z}}_c)\approx(216,227)$ MeV. 
\end{abstract} 
\maketitle
\section{Introduction}
The phase structure of QCD, as a function of temperature $T$ and quark-chemical potential $\mu$, represents the breaking patterns of the relevant symmetries in QCD. Simultaneously, each QCD phase can be characterized by the corresponding order parameters, reflecting the nature of the symmetries. In this sense, exploring the QCD phase diagram is of great importance in understanding strongly interacting systems. Especially, recent energetic progresses, achieved in the ultra-high energy experimental facilities, such as the RHIC, have triggered much interest to investigate the QCD phase structure in the vicinity of high $T\approx T_c$, whereas $\mu$ remains relatively small, being similar to the early universe.

Starting from the first principle, the lattice QCD (LQCD) simulations must be a promising method to investigate this region ($T\ne0$ and $\mu\approx0$) with less difficulties, such as the sign problem~\cite{Alford:1998sd,Hands:1999md,Fodor:2001au,Maezawa:2007fd,Ali Khan:2000iz,Bornyakov:2004ii}. Many attempts have been also done in various effective field-theoretical approaches~\cite{Fukushima:2003fw,Ratti:2004ra,Ratti:2005jh,Ratti:2006gh,Hansen:2006ee,Sakai:2008py,Kiuchi:2005xu,Vanderheyden:2001gx,Sachs:1991en,BoschiFilho:1992ig,Molodtsov:2007ux,Schafer:1995pz,Schwarz:1999dj}. Among them, interestingly enough, the Polyakov-loop-augmented Nambu-Jona-Lasinio (pNJL) model describes the crossover of the two different QCD order parameters for the chiral and $\mathbb{Z}(N_{c})$ symmetries, represented by the chiral condensate $\langle\bar{q}q\rangle\propto\sigma^{2}$ and the traced Polyakov loop $\langle\phi\rangle\equiv\Phi$, respectively~\cite{Fukushima:2003fw,Ratti:2004ra,Ratti:2005jh,Ratti:2006gh,Hansen:2006ee,Sakai:2008py}. If the $\mathbb{Z}(N_{c})$ symmetry is intact, $\Phi$ becomes zero, indicating the confinement phase. On the contrary, provided that the symmetry is broken spontaneously, one has $\Phi\ne0$ for the deconfinement one.

Instanton model can be also thought as an appropriate framework to be employed for this finite-$T$ subject, considering that it has provided remarkable descriptions so far for various nonperturbative QCD and hadron properties. Note that the instanton solution at finite $T$, being periodic in Euclidean time, i.e., {\it caloron} turned out to be essential for this purpose~\cite{Kraan:1998pm,Lee:1998bb}. Nonetheless for its relevance, its practical application is still under development~\cite{Ilgenfritz:2002qs,GarciaPerez:1999ux,Diakonov:2005qa,Diakonov:2004jn,Diakonov:2008sg,Slizovskiy:2007am}. Confinement properties have been discussed as well with semi-classical objects, such as the meron (a half of regular-gauge instanton), by indicating the area law for the Wilson loop~\cite{Lenz:2003jp,Negele:2004hs}. The caloron with non-trivial holonomy, so called the Kraan-van Baal-Lee-Lu (KvBLL) caloron~\cite{Kraan:1998pm,Lee:1998bb}, was taken into account as a lump of dyons to understand the confinement~\cite{Diakonov:2005qa,Diakonov:2004jn,Slizovskiy:2007am}. 

In the present work, we want to construct an effective thermodynamic potential ($\Omega_\mathrm{eff}$) at finite $T$ with $\mu_{\mathrm{R}}=0$, employing the instanton framework. Our strategy is rather simple and practical as follows:
\begin{itemize}
\item
Using the instanton distribution function at finite $T$ from the caloron solution with trivial holonomy (the Harrington-Shepard caloron)~\cite{Harrington:1976dj,Diakonov:1988my}, we first compute the instanton density and average size of instanton as functions of $T$, resulting in that the instanton effect remains finite even beyond the critical temperature $T_c\sim\Lambda_{\mathrm{QCD}}$. Combining these ingredients, we finally obtain ($\bm{k}$(three momentum), $T$(temperature))-dependent constituent-quark mass $M$, $M_{{\bm{k},T}}$, which plays the most important role in the present approach.

\item 
In constructing $\Omega_\mathrm{eff}$, we take into account a practical way, instead of using the caloron and its quark zero-mode solution: $\Omega_\mathrm{eff}$ is obtained by applying the Matsubara formula to the effective action, which is derived from the usual singular-gauge instanton solution at $T=0$, as done usually in effective models~\cite{Sachs:1991en,BoschiFilho:1992ig,Molodtsov:2007ux,Schafer:1995pz,Schwarz:1999dj,Fukushima:2003fw,Ratti:2004ra,Ratti:2005jh,Ratti:2006gh,Hansen:2006ee,Sakai:2008py}.  
 
 \item
The singular-gauge instanton solution is nothing to do with the confinement~\cite{Negele:2004hs}. On the contrary, it explains the nonperturbative QCD properties very well in terms of the spontaneous breakdown of chiral symmetry (SB$\chi$S). Hence, considering the chiral and $\mathbb{Z}(N_{c})$ symmetries on the same footing as in the pNJL model, we introduce the imaginary quark-chemical potential $\mu_{\mathrm{I}}\equiv A_4$, which corresponds to the uniform color gauge field in the Polyakov gauge. It will be translated later as the traced Polyakov loop $\Phi$, as an order parameter for the spontaneous breakdown of $\mathbb{Z}(N_{c})$ symmetry, i.e., the deconfinement phase transition.
\end{itemize}

Considering all these ingredients, we can write a neat expression for $\Omega_\mathrm{eff}$ as a function of $T$ with $M_{{\bm{k},T}}$.  By solving the saddle-point equations with respect to the mean fields, we can draw the curves for $\sigma^{2}$  and $\Phi$ as functions of $T$. From the numerical calculations, we observe that the first-order deconfinment phase transition in pure-glue QCD is modified to the crossover one, according to the mixing of dynamical quarks and $\Phi$ in $\Omega_\mathrm{eff}$. In contrast, the mixing gives only negligible modification on $\sigma^{2}$. As a result, the crossover of the two different QCD  order parameters is shown by the mixing. It also turns out that $T_{c}$ for the deconfinment phase transition ($T^{\mathbb{Z}}_c$) is lowered by about $10\%$ with $M_{\bm{k},T}$, in comparison to the case with a constant $M$ ($M_{0,0}\approx350$ MeV). From this observation, we can conclude that the nontrivial QCD vacuum contribution and partial chiral restoration play a considerable role even for deconfinment phase transition. The numerical results show that $T^{\chi}_{c}=216$ MeV and $T^{\mathbb{Z}}_{c}=227$ MeV. The discrepancy between them becomes about $10$ MeV, which is rather larger than that computed in the local pNJL model~\cite{Ratti:2004ra,Ratti:2005jh,Ratti:2006gh}. We note that the LQCD simulations provides smaller values than ours, whereas $T^{\mathbb{Z}}_{c}$ is comparable~\cite{Alford:1998sd,Hands:1999md,Fodor:2001au,Maezawa:2007fd,Ali Khan:2000iz,Bornyakov:2004ii}. We also find that $\sigma^{2}$ depends much on the partial chiral restoration.

We organize the present work as follows: We explain briefly the instanton properties and the effective chiral action derived at $T=0$ in Section~II. The instanton distribution function at finite $T$ is introduced, and $M_{{\bm k},T}$ is computed in Section~III. In Section~IV, utilizing $M_{{\bm k},T}$ obtained in the previous Section, we construct an effective thermodynamic potential using the Matsubara formula with $\mu\equiv A_4$, which is translated into the traced Polyakov loop. The numerical results for $\sigma^{2}$ and $\Phi$ as functions of $T$ and related discussions are given in Section~V. The final Section is devoted to summary and conclusion. 
\section{Instanton at zero temperature ($T=0$)}
In this section, we want to explain the instanton framework for vacuum ($T=0$ and $\mu_{\mathrm{R}}=0$) briefly. First, we consider a grand canonical ensemble of dilute instantons, which indicate the tunneling in Euclidean time, as a dominant nonpertubative contribution of the QCD vacuum. In this framework, in principle, we have two phenomenolgical parameters such as the average instanton size $\bar{\rho}_{0}\approx1/3$ fm and inter-instanton distance $\bar{R}_{0}\approx1$ fm~\cite{Diakonov:2002fq}, where the subscript zero indicates vacuum. According to these values, the renomalization scale of this framework can be estimated as $\Lambda\approx1/\bar{\rho}_{0}\approx600$ MeV. It is convenient to start with the quark zero mode in the presence of the instanton background in the chiral limit, assuming that the nonperturbative QCD properties are dominated by the zero mode:
\begin{equation}
\label{eq:DEZM}
(i\rlap{/}{\partial}-\rlap{/}{A}_{I\bar{I}})\Psi^{(0)}_{I\bar{I}}=0,
\end{equation}
where $A_{I\bar{I}}$ stands for a singular-gauge (anti)instanton solution, which can be obtained by minimizing the YM action, considering the self-dual equation and the SO(4) symmetry for the field strength tensor $F_{\mu\nu}$ in Euclidean space:
\begin{eqnarray}
\label{eq:IS}
A^{\alpha}_{I\bar{I}\mu}(x)
=\frac{2\,\bar{\eta}^{\alpha\nu}_{\mu}\bar{\rho}^2_0x_{\nu}}
{x^2(x^2+\bar{\rho}^2_0)},
\end{eqnarray}
where $\eta^{\alpha\nu}_{\nu}$ indicates the 't Hooft symbol~\cite{Diakonov:2002fq}. Provided that the zero mode saturates the low-energy hadron phenomena as mentioned, we can then write a single-quark propagator with the zero-mode solution, $\Psi^{(0)}_{I\bar{I}}$, approximately as
\begin{eqnarray}
\label{eq:QP}
S_{I\bar{I}}=\frac{1}{i\rlap{/}{\partial}-
\rlap{/}{A}_{I\bar{I}}}  \approx S_0- 
\frac{\Psi^{(0)\dagger}_{I\bar{I}} \Psi^{(0)}_{I\bar{I}}}{im},
\end{eqnarray}
where $m$ and $S_0$ denote the small current-quark mass $\approx0$ and free-qaurk propagator $1/(i\rlap{/}{\partial})$. Eq.~(\ref{eq:QP}) can be written alternatively, taking into account the Fourier transform of the zero-mode solution, as follows:
\begin{equation}
\label{eq:prop2}
S=\frac{1}{i\rlap{/}{\partial}+iM(i\partial)},
\end{equation}
where the four momentum-dependent $M$ (constituent-quark mass), $M(i\partial)$ is defined by
\begin{equation}
\label{eq:CQM}
M(i\partial)\equiv M_{i\vec{\partial},i\partial_{4}}
=M_{0,0}F^2(i\partial),
\end{equation}
where we have used the subscripts, representing three momentum and energy separately for convenience. $M_{0,0}$ stands for $M$ at zero momentum transfer at $T=0$. It is worth mentioning that $M_{0,0}$ is a function of $m$ in principle~\cite{Nam:2006ng,Goeke:2007bj}. Moreover, the meson-loop (ML) correction beyond the leading-$N_c$ contributions turns out to be critical for the case with $m\ne0$~\cite{Goeke:2007bj,Nam:2008bq}. However, in the chiral limit as employed in the present work, $M_{0,0}$ can be determined self-consistently without the ML correction, keeping only the leading-$N_c$ contributions. The quark form factor, $F(i\partial)$ can be written in terms of the modified Bessel functions as follows: 
\begin{equation}
\label{eq:FF}
F(i\partial)=2t\left[I_0(t)K_1(t)-I_1(t)K_0(t)-\frac{I_1(t)K_1(t)}{t}\right],
\,\,\,\,t=\frac{|i\rlap{/}{\partial}|\bar{\rho}_0}{2}.
\end{equation}
Approximately, this form factor can be parametrized as follows:
\begin{equation}
\label{eq:FFpara}
F(i\partial)=\frac{2}{2+|i\rlap{/}{\partial}|^2\bar{\rho}^2_0}.
\end{equation}
Hereafter, we will make use of this paramterized form factor for numerical calculations. 

With all the ingredients taken into account so far, one can construct an effective partition function, which brings about the quark propagator given in Eq.~(\ref{eq:QP})~\cite{Diakonov:2002fq}:
\begin{equation}
\label{eq:PF}
\mathcal{Z}_{\rm eff}=\int{d\lambda}\,{D\psi}\,{D\psi^{\dagger}}
\exp\left[\int d^4x\psi^{\dagger}(i\rlap{/}{\partial}+im)\psi
+\lambda(Y^++Y^-)
+N(\ln\frac{N}{\lambda V\mathcal{M}}-1)\right],
\end{equation}
where we have introduced $\lambda$ as a Lagrangian multiplier to make the partition function exponent and an arbitrary massive parameter $\mathcal{M}$ to make the argument of the logarithm dimensionless. It is assumed that the numbers of the instanton and anti-instanton are the same, $N_I=N_{\bar{I}}=N/2$ indicating the $CP$-invariant vacuum, so that $N$ denotes the sum of the total number of the pseudo-particles. Since we are interested in the case for $N_f=2$, the $2N_f$-'t Hooft interaction $Y^{\pm}$ can be casted into a four-quark instanton-induced interaction. From this Nambu-Jona-Lasinio (NJL) model-like four-quark interaction, one can construct an effective chiral action as a functional of quarks, and scalar and pseudo-scalar mesons, by an exact bosonization process~\cite{Diakonov:2002fq}. After freezing out all the meson fields, except for the isosinglet scalar meson ($\sigma$), which is responsible for the spontaneous breakdown of chiral symmetry (SB$\chi$S), we have the following effective chiral action:
\begin{equation}
\label{eq:ECA}
\mathcal{S}_\mathrm{eff}=N\left[1-\ln\frac{N}{\lambda V\mathcal{M}}\right]
+2V\sigma^{2}
-2VN_cN_f\int\frac{d^4k}{(2\pi)^4}
\ln\left[k^2+M^2_{0,0}F^4(k^2)\right],
\end{equation}
where $\mathcal{S}_\mathrm{eff}$ has been written in momentum space. More detailed explanations on the derivation of the effective action in Eq.~(\ref{eq:ECA}) from the partition function in Eq.~(\ref{eq:PF}) can be found in  Refs.~\cite{Diakonov:2002fq,Goeke:2007bj,Nam:2008bq}. $M_{0,0}$, as a function of the saddle-point value of $\sigma$ ($\sigma_0$) and $\rho_0$, can be computed with the normalized instanton distribution function $d_N(\rho)$, which will be discussed in detail in the next section: 
\begin{equation}
\label{eq:COMSM}
M_{0,0}=\mathrm{M}\,\sigma_0
\int\,d\rho\,d_N(\rho)(2\pi\rho)^2
=\mathrm{M}\,\sigma_0(2\pi\bar{\rho}_0)^2.
\end{equation}
Here, we have assumed that $d_N(\rho)\approx\delta(\rho-\bar{\rho}_{0})$ in the large-$N_c$ limit~\cite{Diakonov:1995qy}, where $\mathrm{M}$ stands for a massive quantity, $\sqrt{\lambda}/(2N^{2}_{c})$. Considering the saddle-point (self-consistent) equations with respect to $\lambda$ and $\sigma$, one is led to the following relations:
\begin{eqnarray}
\label{eq:SPE1}
\frac{\delta\mathcal{S}_\mathrm{eff}}{\delta\lambda}&=&0\to
\frac{N}{V}=4N_c\int\frac{d^4k}{(2\pi)^4}
\frac{M^2_{0,0}F^4(k^2)}{k^2+M^2_{0,0}F^4(k^2)}\equiv{n_{0}},
\\
\label{eq:SPE2}
\frac{\delta\mathcal{S}_\mathrm{eff}}{\delta\sigma}&=&0\to
\sigma^2_0=2N_c\int\frac{d^4k}{(2\pi)^4}
\frac{M^2_{0,0}F^4(k^2)}{k^2+M^2_{0,0}F^4(k^2)},
\end{eqnarray}
where we assigned the instanton number density (packing fraction) $N/V$ as $n_{0}$. Considering the dilute-instanton ensemble, $n_{0}\approx(200\,\mathrm{MeV})^4$, $\bar{\rho}_{0}\approx0.3$ fm, and $\bar{R}_{0}\approx1.0$ fm, and substituting Eq.~(\ref{eq:CQM}) to Eq.~(\ref{eq:SPE1}), we can determine $M_{0,0}\approx350$ MeV self-consistently. Moreover, by equating Eqs.~(\ref{eq:SPE1}) and (\ref{eq:SPE2}), we have a relation $2\sigma^2_0=n_{0}$ at the saddle point. Using this relation, Eq.~(\ref{eq:COMSM}) can be rewritten as 
\begin{equation}
\label{eq:COMSM1}
M_{0,0}=\mathrm{M}\sqrt{2n_{0}}(2\pi\bar{\rho}_0)^{2}
\approx350\,\mathrm{MeV}.
\end{equation}
Note that, in medium, $\sigma_0$ plays a role of an order parameter for the chiral phase transition as will be discussed in the following Sections, being similar to the chiral condensate $\langle iq^\dagger q\rangle$, which reads:
\begin{equation}
\label{eq:CCC}
\langle iq^\dagger q\rangle=4N_c\int\frac{d^4k}{(2\pi)^4}
\frac{M_{0,0}F^2(k^2)}{k^2+M^2_{0,0}F^4(k^2)}.
\end{equation}

\section{Instanton at finite temperature ($T\ne0$)}
Now, we want to discuss the instanton at finite $T$ in the mean-field approach. In Ref.~\cite{Kiuchi:2005xu}, it was assumed that the relevant quantities such as $n_{0}$, $\bar{\rho}_{0}$, and $M_{0,0}$ remain the same with their vacuum values below $T_c$ and disappear beyond it. However, more realistic situation near $T_c$ is quite complicated as indicated in Refs.~\cite{Diakonov:1988my,Schafer:1996wv}, in which the temporally periodic instantons with the trivial holonomy, i.e., the Harrington-Shepard caloron was employed~\cite{Harrington:1976dj}. These works showed that the relevant quantities decrease with respect to $T$ smoothly, and they beomce finite even beyond $T_c$. 

\begin{figure}[t]
\includegraphics[width=8cm]{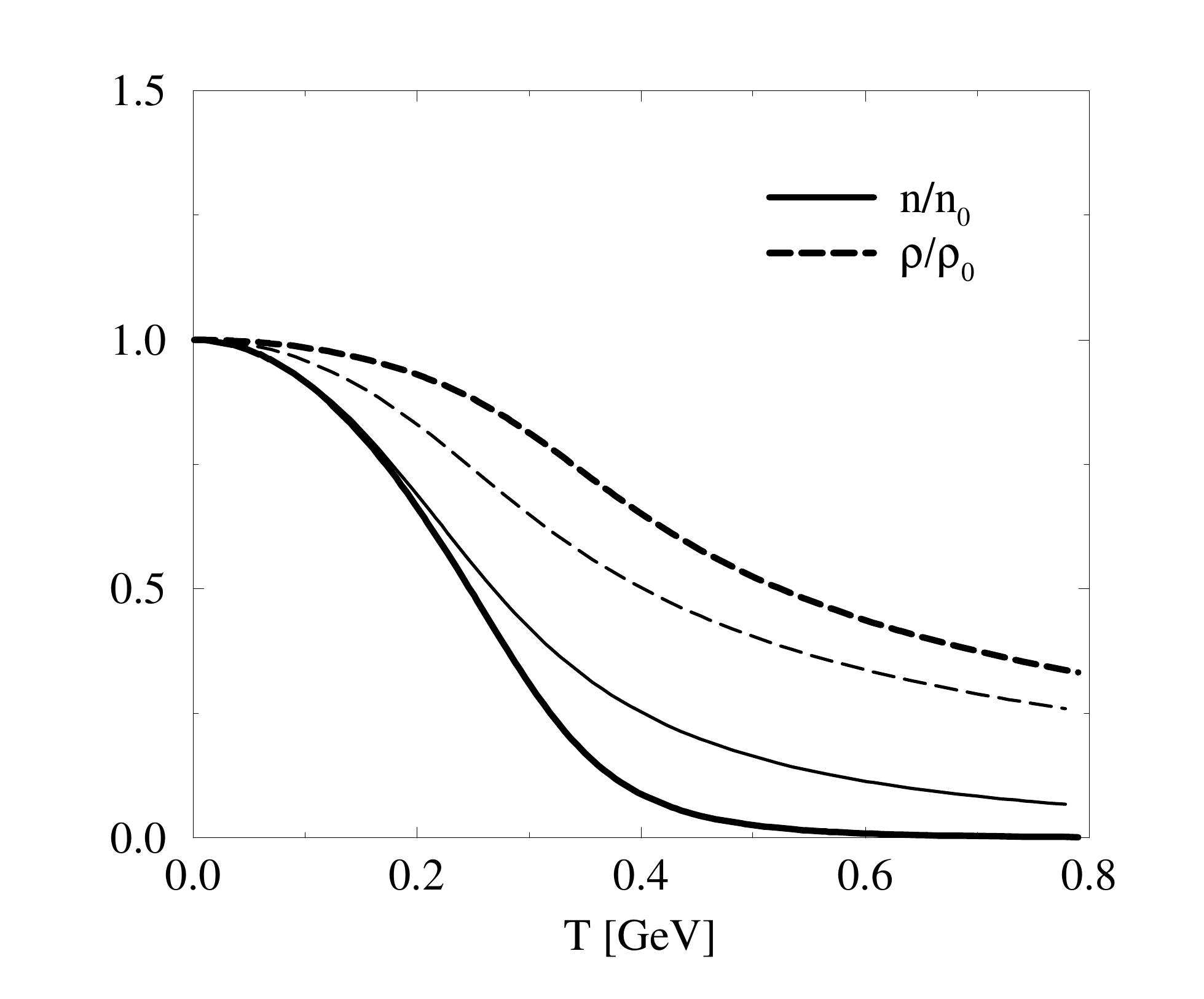}
\caption{$n/n_{0}$ (solid) and $\bar{\rho}/\bar{\rho}_0$ (dashed) as functions of d$T$ for $N_c=3$ (thick lines) and $2$ (thin lines). Here, we use $n_{0}=(0.2\,\mathrm{GeV})^4$ and $\bar{\rho}=(0.6\,\mathrm{MeV})^{-1}$.}       
\label{fig0}
\end{figure}

We are in a position to consider the $T$-dependence of these relevant quantities, $n$, $\bar{\rho}$, and $M$ using the Harrington-Shepard caloron in detail. An instanton distribution function for arbitrary $N_c$ and $N_f$ can be expressed with a Gaussian suppression factor as a function of $T$ and an arbitrary instanton size $\rho$, employing the variational method for pure-glue QCD~\cite{Diakonov:1988my}:
\begin{equation}
\label{eq:IND}
d(\rho,T)=\underbrace{C_{N_c}\,\Lambda^b_{\mathrm{RS}}\,
\hat{\beta}^{N_c}}_\mathcal{C}\,\rho^{b-5}
\exp\left[-(A_{N_c}T^2
+\bar{\beta}\gamma n\bar{\rho}^2)\rho^2 \right].
\end{equation}
Here, we considered the $CP$-invariant vacuum as before and assumed the same analytical form of the distribution function for both the insanton and anti-instanton. Note that $n$ and $\bar{\rho}$ have been taken into account as functions of $T$ implicitly. We also assigned the constant factor in the {\it r.h.s.} of the above equation as $\mathcal{C}$ for simplicity. The abbreviated notations are also given as:
\begin{equation}
\label{eq:para}
\hat{\beta}=-b\ln[\Lambda_\mathrm{RS}\rho_\mathrm{cut}],\,\,\,\,
\bar{\beta}=-b\ln[\Lambda_\mathrm{RS}\langle R\rangle],\,\,\,
C_{N_c}=\frac{4.60\,e^{-1.68\alpha_{\mathrm{RS}} Nc}}{\pi^2(N_c-2)!(N_c-1)!},
\end{equation}
\begin{equation}
\label{eq:AA}
A_{N_c}=\frac{1}{3}\left[\frac{11}{6}N_c-1\right]\pi^2,\,\,\,\,
\gamma=\frac{27}{4}\left[\frac{N_c}{N^2_c-1}\right]\pi^2,\,\,\,\,
b=\frac{11N_c-2N_f}{3}.
\end{equation}
Note that we defined the one-loop inverse charge $\hat{\beta}$ and $\bar{\beta}$ at certain phenomenological cutoff values $\rho_\mathrm{cut}$ and $\langle R\rangle\approx\bar{R}$. As will be shown, only $\bar{\beta}$ is relevant in the following discussions and will be fixed self-consistently within the present framework. $\Lambda_{\mathrm{RS}}$ stands for a scale, depending on a renormalization scheme, whereas $V_3$ for the three-dimensional volume. Using the instanton distribution function in Eq.~(\ref{eq:IND}), we can compute the average value of the instanton size, $\bar{\rho}^2$ straightforwardly as follows~\cite{Schafer:1996wv}:
\begin{equation}
\label{eq:rho}
\bar{\rho}^2(T)
=\frac{\int d\rho\,\rho^2 d(\rho,T)}{\int d\rho\,d(\rho,T)}
=\frac{\left[A^2_{N_c}T^4
+4\nu\bar{\beta}\gamma n \right]^{\frac{1}{2}}
-A_{N_c}T^2}{2\bar{\beta}\gamma n},
\end{equation}
where $\nu=(b-4)/2$. It can be easliy shown that Eq.~(\ref{eq:rho}) satisfies the  following asymptotic behaviors~\cite{Schafer:1996wv}:
\begin{equation}
\label{eq:asym}
\lim_{T\to0}\bar{\rho}^2(T)=\sqrt{\frac{\nu}{\bar{\beta}\gamma n}},
\,\,\,\,
\lim_{T\to\infty}\bar{\rho}^2(T)=\frac{\nu}{A_{N_c}T^2}.
\end{equation}
Note that the second relation of Eq.~(\ref{eq:asym}) provides a correct scale-temperature behavior at high $T$, i.e., $1/\bar{\rho}\approx\Lambda\propto T$. Substituting Eq.~(\ref{eq:rho}) into Eq.~(\ref{eq:IND}), the distribution function can be evaluated further as:
\begin{equation}
\label{eq:dT}
d(\rho,T)=\mathcal{C}\,\rho^{b-5}
\exp\left[-\mathcal{F}(T)\rho^2 \right],\,\,\,\,
\mathcal{F}(T)=\frac{1}{2}A_{N_c}T^2+\left[\frac{1}{4}A^2_{N_c}T^4
+\nu\bar{\beta}\gamma n \right]^{\frac{1}{2}}.
\end{equation}
The instanton density $n$ can be computed self-consistently, using the following equation:
\begin{equation}
\label{eq:NOVV}
n^\frac{1}{\nu}\mathcal{F}(T)=\left[\mathcal{C}\,\Gamma(\nu) \right]^\frac{1}{\nu},
\end{equation}
where we have replaced $NT/V_3\to n$, and $\Gamma(\nu)$ indicates the $\Gamma$-fucntion with an argument $\nu$. Note that $\mathcal{C}$ and $\bar{\beta}$ can be determined easily using Eqs.~(\ref{eq:rho}) and (\ref{eq:NOVV}), incorporating the vacuum values for $n$ and $\bar{\rho}$: $\mathcal{C}\approx9.81\times10^{-4}$ and $\bar{\beta}\approx9.19$.  The numerical results for the normalized $n/n_0$ (solid) and $\bar{\rho}/\bar{\rho}_0$ (dashed) are shown as functions of $T$ in Figure~\ref{fig0} for $N_c=3$ (thick lines) and $2$ (thin lines), separately. As shown in the figure, although $n$ and $\bar{\rho}$ get decreased gradually as $T$ increases, indicating the reduction of the instanton contributions, they remain finite even beyond $T_c\sim\Lambda_{\mathrm{QCD}}\approx200$ MeV. 

\begin{figure}[t]
\includegraphics[width=8cm]{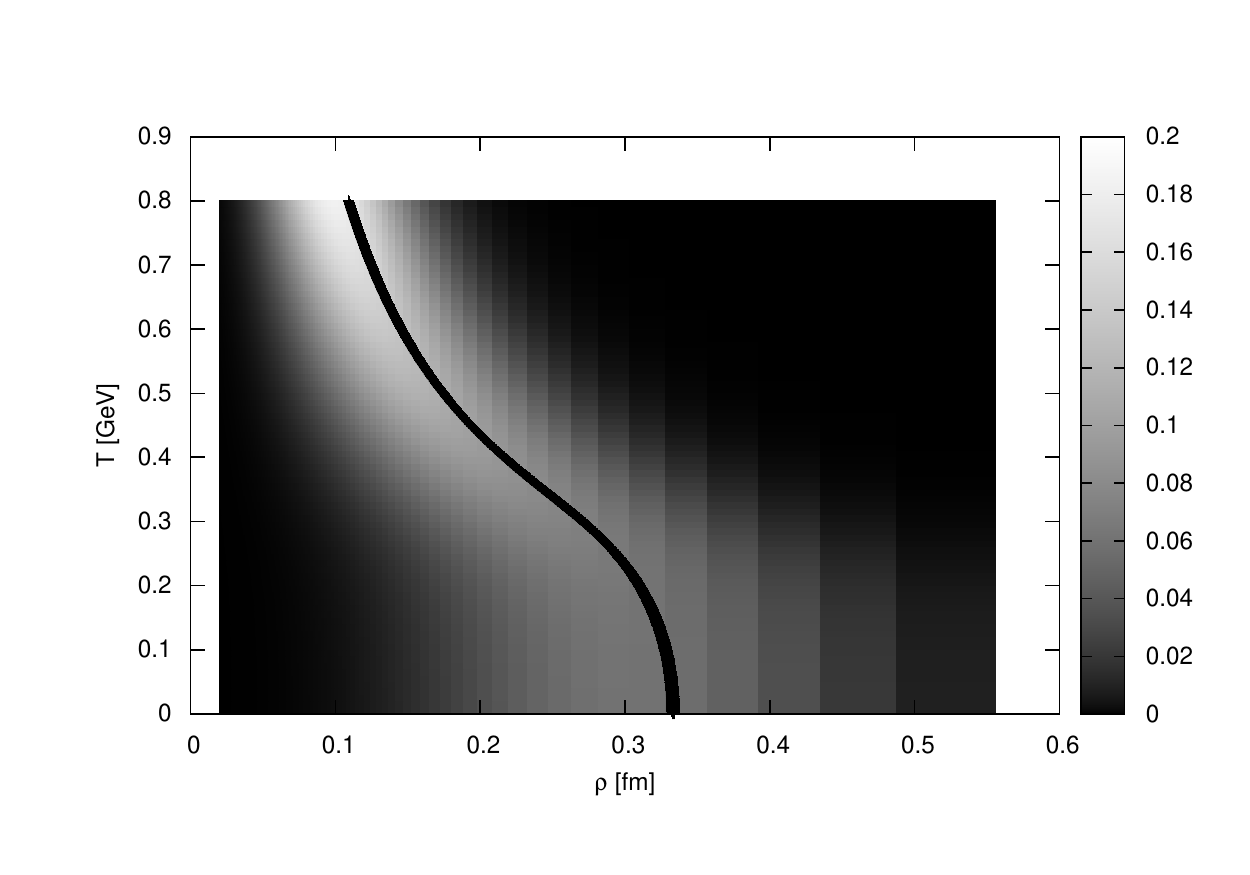}
\caption{$d_N(\rho, T)$ on the $\rho-T$ plane. The trajectory on the curve represents $\bar{\rho}(T)$ in Eq.~(\ref{eq:rho}).}       
\label{fig1}
\end{figure}

Finally, in order to estimate the $T$-dependence of $M$, it is necessary to consider the normalized distribution function (see Eq.~(\ref{eq:COMSM})), defined as follows:
\begin{equation}
\label{eq:NID}
d_N(\rho,T)=\frac{d(\rho,T)}{\int d\rho\,d(\rho,T)}
=\frac{\rho^{b-5}\mathcal{F}^\nu(T)
\exp\left[-\mathcal{F}(T)\rho^2 \right]}{\Gamma(\nu)}.
\end{equation}
The numerical result for $d_N(\rho,T)$ is given in Figure~\ref{fig1}. As shown in the figure, the peak of the $d_N(\rho,T)$ surface moves to smaller $\rho$ as $T$ increases on the $\rho$-$T$ plane. Now, we want to employ the large-$N_c$ limit to simplify the expression of $d_N(\rho,T)$ as done for the $T=0$ case in the previous Section. Since the  parameter $b$ is in the order of $\mathcal{O}(N_c)$ as shown in Eq.~(\ref{eq:para}), it becomes infinity as $N_c\to\infty$, and the same for $\nu$. In this limit, as understood from Eq.~(\ref{eq:NID}), $d_N(\rho,T)$ can be approximated as a $\delta$-function~\cite{Diakonov:1995qy}: 
\begin{equation}
\label{eq:NID2}
\lim_{N_c\to\infty}d_N(\rho,T)=\delta({\rho-\bar{\rho}\,(T)}).
\end{equation}
The trajectory of this $\delta$-function projected on the $\rho$-$T$ plane is depicetd in Figure~\ref{fig1} by the thick-solid line. Substituting Eq.~(\ref{eq:NID2}) into Eq.~(\ref{eq:COMSM}) and using Eq.~(\ref{eq:FFpara}), we can write the ($\bm{k},T$)-dependent $M$ as follows:
\begin{equation}
\label{eq:momo}
M_{\bm{k},k_{4}}=
\underbrace{M_{0,0}\left[\frac{\sqrt{n}\,\bar{\rho}^2}
{\sqrt{n_{0}}\,\bar{\rho}^2_0}\right]}_{M_{0,T}}
\left[\frac{2}{2+(\bm{k}^2+k^2_4)\,\bar{\rho}^2} \right]^2,
\end{equation}
where we use $M_{0,0}\approx350$ MeV, considering that Eq.~(\ref{eq:momo}) must be identical to Eq.~(\ref{eq:CQM}) as $T\to0$. Moreover, we assumed that the Lagrangian multiplier $\lambda$, equivalently $\mathrm{M}$, is independent on $T$. Note that the fourth component of the momentum, $k_4$ will be replaced as a function of $T$ by employing the Matsubara frequency in the next Section. 

Here, we discuss a caveat in our model. In deriving the momentum- and temperature-dependent constituent-quark mass in Eq.~(\ref{eq:momo}), we have used a fact that the Harrington-Shepard caloron is the zero-mode solution of the Dirac equation with the trivial holonomy in the presence of the instanton background at finite $T$. However, as mentioned in Section I, we will employ the Polyakov loop in Section IV, corresponding to the nontrivial holonomy, resulting in that the Harrington-Shepard caloron is not the exact zero-mode solution of the Dirac equation anymore. Instead, the KvBLL caloron becomes the exact one. 

Since the remedy for this problematic issue is quite complicated and beyond our scope in the present work, we assume that the Harrington-Shepard caloron is almost the zero-mode solution even with the nontrivial holonomy so that Eq.~(\ref{eq:momo}) is qualitatively applicable.

\section{Effective thermodynamic potential}
\begin{figure}[t]
\includegraphics[width=8cm]{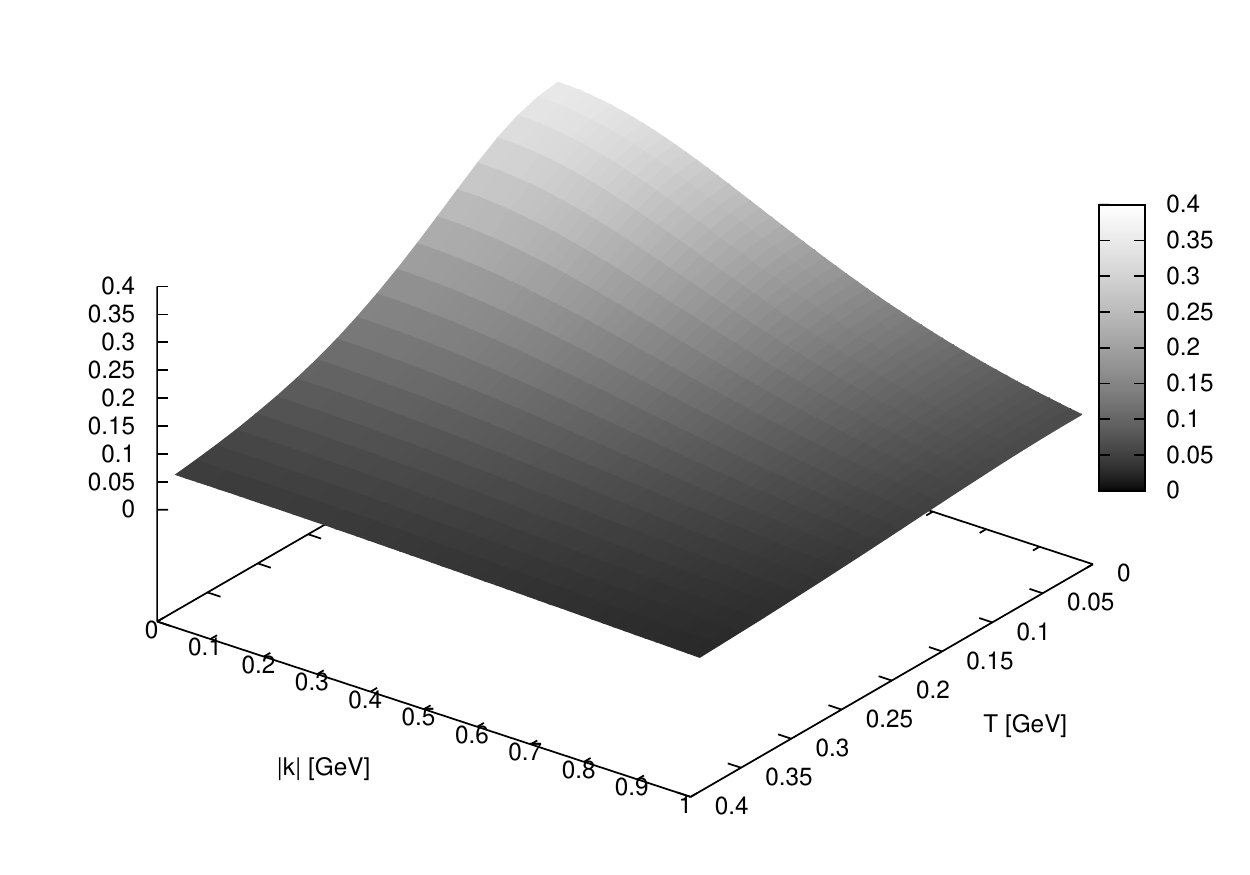}
\caption{$M_{\bm{k},T}$ in Eq.~(\ref{eq:ENERGY}) as a function of $T$ and the absolute alue of three momentum $|\bm{k}|$ [GeV].}       
\label{fig2}
\end{figure}

In this section, we construct an effective thermodynamic potential $\Omega_\mathrm{eff}$, considering all the ingredients discussed so far. In addition, we take into account the imaginary quark chemical potential ($\mu_{\mathrm{I}}\equiv A_{4}$), which will be translated as the traced Polyakov loop as an order parameter for the deconfinment phase transition~\cite{Fukushima:2003fw,Sakai:2008py}. Moreover, this corresponds to the uniform color gauge field, induced in the Polyakov gauge, $A_{\mu}=(\vec{0},A_{4})$. All calculations will be performed in the case for $N_c=3$, $N_f=2$, and $\mu_{\mathrm{R}}=0$ in the chiral limit for the leading-$N_c$ contributions. In order to evaluate $\Omega_\mathrm{eff}$ as a function of $T$ from the effective action in Eq.~(\ref{eq:ECA}), we employ the anti-periodic Matsubara formula for fermions in Euclidean space: 
\begin{equation}
\label{eq:MTBR}
\int\frac{d^4k}{(2\pi)^4}\,f(\bm{ k},k_4)\to T\sum^{\infty}_{n=-\infty}
\int\frac{d^3\bm{ k}}{(2\pi)^3}\,f(\bm{k},w_{n}),
\end{equation}
where the Matsubara frequency reads $w_n=(2n+1)\pi T$. We also include an imaginary quark-chemical potential~\cite{Fukushima:2003fw,Ratti:2006gh}, which can be identified as the fourth component of SU($N_c$) gauge field ($A_4$) in Euclidean space, resulting in a replacement $k\to k-A_4$ in $\Omega_\mathrm{eff}$. Applying these ingredients to Eq.~(\ref{eq:ECA}), we can have the following effective thermodynamic potential per unit volume with the mixing of the dynamical quarks and $A_{4}$:
\begin{equation}
\label{eq:TDPLP}
\Omega^\mathrm{q+A_{4}}_\mathrm{eff}=
2\sigma^{2}-N_fT\sum^{\infty}_{n=-\infty}
\int\frac{d^3\bm{k}}{(2\pi)^3}
\mathrm{Tr}_c\ln\left[\frac{(k-A_4)^2+M^2_{\bm{k},T}}{T^2} \right],
\end{equation}
where $\sigma$ is a function of $T$ as well. Note that we have ignored $A_{4}$ in $M$ for simplicity. After a straightforward manipulation, using the relations and identities given in Appendix, one is led to the following expression:
\begin{equation}
\label{eq:TDPLP2}
\Omega^\mathrm{q+A_{4}}_\mathrm{eff}\approx2\sigma^2-N_fT
\int\frac{d^3\bm{k}}{(2\pi)^3}\mathrm{Tr}_c
\left[\frac{E_{\bm{k},T}}{T}
+\ln\left[\left(1+e^{-\frac{E_{\bm{k},T}-iA_4}{T}}\right)
\left(1+e^{-\frac{E_{\bm{k},T}+iA_4}{T}}\right)\right]\right].
\end{equation}
Here, we have used that
\begin{equation}
\label{eq:ENERGY}
E_{\bm{k},T}=\left(\bm{k}^2+M^2_{\bm{k},T}\right)^{1/2},\,\,\,\,
M_{\bm{k},T}=M_{0,T}\left[\frac{2}{2+\bm{k}^2\bar{\rho}^2}\right]^2,
\end{equation}
where $M_{0,T}$ is defined in Eq.~(\ref{eq:momo}). For more details, see also Appendix. The numerical result for $M_{\bm{k},T}$ is shown in Figure~\ref{fig2} as a function of $T$ and $|\bm{k}|$. From the figure, it turns out that the $\bm{k}$ dependence of $M$ becomes weak as $T$ increases. At $T=0.4$ GeV, the $\bm{k}$ dependence does not appear at all. This behavior can be understood as follows: The $\bm{k}$ dependence of $M$ is generated  from the quark-instanton interaction, i.e, the delocalization of the quark fields in the instanton ensemble. As the instanton ensemble becomes more dilute as $T$ increases (smaller $\bar{\rho}$, equivalently), the interaction probability decreases, resulting in the reduction of the $\bm{k}$ dependence. However, note that the dependence still remains visible around $T_c\sim\Lambda_{\mathrm{QCD}}\approx200$ MeV as shown in the figure.

Now, we are in a position to consider the traced Polyakov loop $\phi$, defined in a SU($N_c$) gauge group as:
\begin{equation}
\label{eq:POL}
\phi=\frac{1}{N_c}\mathrm{Tr}_c\exp\left(\frac{iA_4}{T}\right),
\,\,\,\,
\phi^*=\frac{1}{N_c}\mathrm{Tr}_c\exp\left(\frac{-iA_4}{T}\right).
\end{equation}
Taking into account the Polyakov gauge, $A_4$ is diagonal in a $N_c\times N_c$ matrix. For instance, the perturbative YM potential can be expressed by this physical quantity~\cite{Gross:1980br} and prefers the deconfinment phase~\cite{Polyakov:1976fu,Polyakov:1978vu}: $\langle\phi\rangle$ becomes finite according to the spontaneous breakdown of the $Z_{N_c}$ symmetry with the trivial holonomy. In contrast, if the symmetry is intact, one finds $\langle\phi\rangle=0$, indicating the confining phase with the non-trivial holonomy.

Thus, $\phi$ plays the role of an exact order parameter for the $\mathbb{Z}(N_{c})$ symmetry for pure-glue QCD, in which the quark degree of freedom is decoupled according to its infinitely heavy mass. It is worth mentioning that the $\mathbb{Z}(N_{c})$ symmetry is broken explicitly in the presence of dynamical quarks with finite mass, considering the anti-symmetric nature of fermions, resulting in that $\langle\phi\rangle$ is not an exact order parameter for $Z_{N_c}$ symmetry any more. However, from the phenomenological point of view, incorporating $\phi$ and dynamical quarks has been quite successful to a certain extent to explain various features of the QCD phases transition: The crossover near $T_c$ for instance for $N_{f}=2$. Hence, in the present work as done in the pNJL model, we have incorporated the instanton-based model with the SB$\chi$S and the traced Polyakov loop as an order parameter for the $\mathbb{Z}(N_{c})$ symmetry. The trace over color space in Eq.~(\ref{eq:TDPLP2}) can be evaluated further in terms of $\phi$ and $\phi^*$ using Eq.~(\ref{eq:POL}) as follows:
\begin{eqnarray}
\label{eq:TRR}
&&\mathrm{Tr}_c\ln\left[\left(1+e^{-\frac{E_{\bm{k},T}-iA_4}{T}}\right)
\left(1+e^{-\frac{E_{\bm{k},T}+iA_4}{T}}\right)\right]
\nonumber\\
&&=\ln\left[1+N_c\left(\phi+\phi^*\,
e^{-\frac{E_{\bm{k},T}}{T}}\right)e^{-\frac{E_{\bm{k},T}}{T}}
+e^{-\frac{3E_{\bm{k},T}}{T}}\right]
\nonumber\\
&&+\ln\left[1+N_c\left(\phi^*+\phi\,
e^{-\frac{E_{\bm{k},T}}{T}}\right)e^{-\frac{E_{\bm{k},T}}{T}}
+e^{-\frac{3E_{\bm{k},T}}{T}}\right].
\end{eqnarray}

On top of $\Omega^\mathrm{q+A_{4}}_\mathrm{eff}$ (now becoming $\Omega^\mathrm{q+\phi}_\mathrm{eff}$), an additional pure-glue effective thermodynamic potential was suggested in Refs.~\cite{Ratti:2004ra,Ratti:2005jh,Ratti:2006gh,Hansen:2006ee,Sakai:2008py} as a function of $\phi$ and $\phi^*$:
\begin{equation}
\label{eq:POLPO}
\Omega^\mathrm{\phi}_\mathrm{eff}=
-T^4\left[\frac{b_2(T)}{2}(\phi\,\phi^*)
+\frac{b_3}{6}(\phi^3+\phi^{*3})-\frac{b_4}{4}(\phi\,\phi^*)^2 \right],
\end{equation}
where the coefficient $b_2$ is a function of $T$:
\begin{eqnarray}
\label{eq:PLP}
b_2(T)&=&a_0+a_1\left[\frac{T_0}{T}\right]
+a_2\left[\frac{T_0}{T}\right]^2
+a_3\left[\frac{T_0}{T}\right]^3.
\end{eqnarray}
Here, $T_0$ denotes critical $T$ in pure-glue QCD, resulting in $T_0=270$ MeV at which the first-order phase transition occurs. Note that $T_{0}$ is different from $T_{c}$, which will be computed later in the presence of the mixing of the dynamical quarks and $\phi$. The coefficients, $a$ and $b$, are listed in Table~\ref{table1}~\cite{Boyd:1996bx}. This parameterization of the effective potential in Eq.~(\ref{eq:POLPO}) bears the $Z_{N_c}$ symmetry, conserved in pure-glue QCD by construction, and works qualitatively well up to $T\approx(2-3)\,T_c$, from which the transverse gluons come into play considerably. 
\begin{table}[b]
\begin{tabular}{c|c|c|c|c|c}
$\hspace{0.4cm}a_0\hspace{0.4cm}$
&$\hspace{0.4cm}a_1\hspace{0.4cm}$
&$\hspace{0.4cm}a_2\hspace{0.4cm}$
&$\hspace{0.4cm}a_4\hspace{0.4cm}$
&$\hspace{0.4cm}b_3\hspace{0.4cm}$
&$\hspace{0.4cm}b_4\hspace{0.4cm}$\\
\hline
$6.75$&$-1.95$&$2.63$&$-7.44$&$0.75$&$7.50$
\end{tabular}
\caption{Coefficients for $\Omega^\mathrm{\phi}_\mathrm{eff}$ in Eqs.~(\ref{eq:POLPO}) and (\ref{eq:PLP}), taken from Refs.~\cite{Ratti:2004ra,Ratti:2005jh,Ratti:2006gh}.}
\label{table1}
\end{table}

Finally, substituting Eq.~(\ref{eq:TRR}) into Eq.~(\ref{eq:TDPLP2}) and adding Eq.~(\ref{eq:POLPO}) to it, we arrive at the following expression for the effective thermodynamic potential with the two order parameters, $\sigma^{2}$ for the chiral phase and $\phi$ $(\phi^{*})$ for the deconfinment phase transitions, at finite $T$ and $\mu_{R}=0$: 
\begin{eqnarray}
\label{eq:TDPLP3}
\Omega_\mathrm{eff}
&=&\Omega^\mathrm{q+\Phi}_\mathrm{eff}
+\Omega^\mathrm{\Phi}_\mathrm{eff}
=2\sigma^2-2N_f\Bigg[N_c
\int\frac{d^3\bm{k}}{(2\pi)^3}E_{\bm{k},T}
\nonumber\\
&+&T\int\frac{d^3\bm{k}}{(2\pi)^3}
\ln\left[1+N_c\left(\Phi+\bar{\Phi}\,e^{-\frac{E_{\bm{k},T}}{T}}\right)
e^{-\frac{E_{\bm{k},T}}{T}}+e^{-\frac{3E_{\bm{k},T}}{T}}\right]
\nonumber\\
&+&T\int\frac{d^3\bm{k}}{(2\pi)^3}
\ln\left[1+N_c\left(\bar{\Phi}+\Phi\,e^{-\frac{E_{\bm{k},T}}{T}}\right)
e^{-\frac{E_{\bm{k},T}}{T}}+e^{-\frac{3E_{\bm{k},T}}{T}}\right]\Bigg]
\nonumber\\
&-&T^4\left[\frac{b_2(T)}{2}(\Phi\,\bar{\Phi})
+\frac{b_3}{6}(\Phi^3+\bar{\Phi}^3)-\frac{b_4}{4}(\Phi\,\bar{\Phi})^2 \right],
\end{eqnarray}
where we have replaced $\phi$ into its mean value $\langle\phi\rangle\equiv\Phi$ as done in Refs.~\cite{Ratti:2004ra,Ratti:2005jh,Ratti:2006gh}. Although, this expression for $\Omega_\mathrm{eff}$ is very similar to those given in Refs.~\cite{Ratti:2004ra,Ratti:2005jh,Ratti:2006gh,Hansen:2006ee,Sakai:2008py}, ours is distinctive from them quantitatively in several points:
\begin{itemize}
\item
The scale parameter of the model $\Lambda\approx1/\bar{\rho}$ is obtained as a function of $T$ by solving the instanton distribution function as discussed in the previous Section. Moreover, $M$ is expressed as a function of $\bm{k}$  and $T$ ($M_{\bm{k},T}$), rather than a constant, manifesting the partial chiral restoration and nontrivial QCD vacuum contributions.
\item
Consequently, there appears no divergence in the energy integral $\propto\int d^3\bm{k}\,E_{\bm{k},T}$ in Eq.~(\ref{eq:TDPLP3}) by virtue of the $\bm{k}$-dependent $M$, which plays the role of an intrinsic ultraviolet (UV) regulator, being different from other local-interaction models, such as the usual pNJL model.  
\item
At the smae time, as  for the $2N_f$-'t Hooft interaction, the quark-meson coupling strength depends on $\bm{k}$ as well as $T$, being different from that in other models, in which it is a constant value fixed at zero $T$. 
\item
All the relevant quantities at zero $T$, $M_{0,0}$, $\sigma_0$ and $n_{0}$ for instance, are determined self-consistently by solving the saddle-point equations, Eqs.~(\ref{eq:SPE1}) and (\ref{eq:SPE2}).
\end{itemize}

Now, we evalute the equations of motion with respect to the mean fields, $\sigma$, $\Phi$, and $\bar{\Phi}$, by minimizing $\Omega_\mathrm{eff}$:
\begin{equation}
\label{eq:SD}
\frac{\delta\Omega_\mathrm{eff}}{\delta\Phi}=0,\,\,\,\,
\frac{\delta\Omega_\mathrm{eff}}{\delta\bar{\Phi}}=0,\,\,\,\,
\frac{\delta\Omega_\mathrm{eff}}{\delta\sigma}=0.
\end{equation}
From the first two equations, it can be easily seen that $\Phi=\bar{\Phi}$ at the saddle point, and we have the relation as follows:
\begin{equation}
\label{eq:WRP}
T^3\left[b_4\,\Phi^3-b_3\,\Phi^2-b_2(T)\,\Phi\right]=
\int\frac{d^3\bm{k}}{(2\pi)^3}
\frac{4N_cN_f\left(e^{-\frac{E_{\bm{k},T}}{T}}
+e^{-2\frac{E_{\bm{k},T}}{T}}\right)}{1+N_c\Phi
\left(1+e^{-\frac{E_{\bm{k},T}}{T}}\right)
e^{-\frac{E_{\bm{k},T}}{T}}+e^{-\frac{3E_{\bm{k},T}}{T}}}.
\end{equation}
Similarly, the last equation in Eq.~(\ref{eq:SD}) gives
\begin{equation}
\label{eq:WRS}
2\sigma^2=N_cN_f\int\frac{d^3\bm{k}}{(2\pi)^3}
\frac{M^2_{\bm{k},T}}{E_{\bm{k},T}}
\left[1-\frac{2\left(\Phi\,e^{-\frac{E_{\bm{k},T}}{T}}+2\,\Phi\,
e^{-\frac{2E_{\bm{k},T}}{T}}+e^{-\frac{3E_{\bm{k},T}}{T}}\right)}
{1+N_c\Phi\left(1+e^{-\frac{E_{\bm{k},T}}{T}}\right)
e^{-\frac{E_{\bm{k},T}}{T}}
+e^{-\frac{3E_{\bm{k},T}}{T}}} \right],
\end{equation}
where we have used the following relation 
\begin{equation}
\label{eq:rel}
\frac{\partial E_{\bm{k},T}}{\partial\sigma}
=\frac{1}{\sigma}\frac{M^2_{\bm{k},T}}{E_{\bm{k},T}},
\end{equation}
considering Eqs.~(\ref{eq:COMSM}) and (\ref{eq:COMSM1}). From Eq.~(\ref{eq:WRS}), one can immediately see the following asymptotic behaviors:
\begin{eqnarray}
\label{eq:ASY}
\lim_{T\to0}[2\sigma^2]&=&N_cN_f\int\frac{d^3\bm{k}}{(2\pi)^3}
\frac{M^2_{\bm{k},T}}{E_{\bm{k},T}}\,\,\,\,\mathrm{as}\,\,\,\,
\Phi\to0, 
\nonumber\\
\lim_{T\to\infty}[2\sigma^2]&=&0\,\,\,\,\mathrm{as}\,\,\,\,
\Phi\to1,
\end{eqnarray}
showing appropriate chiral properties, as expected.

\section{Numerical results and Discussions}
\begin{figure}[t]
\includegraphics[width=8cm]{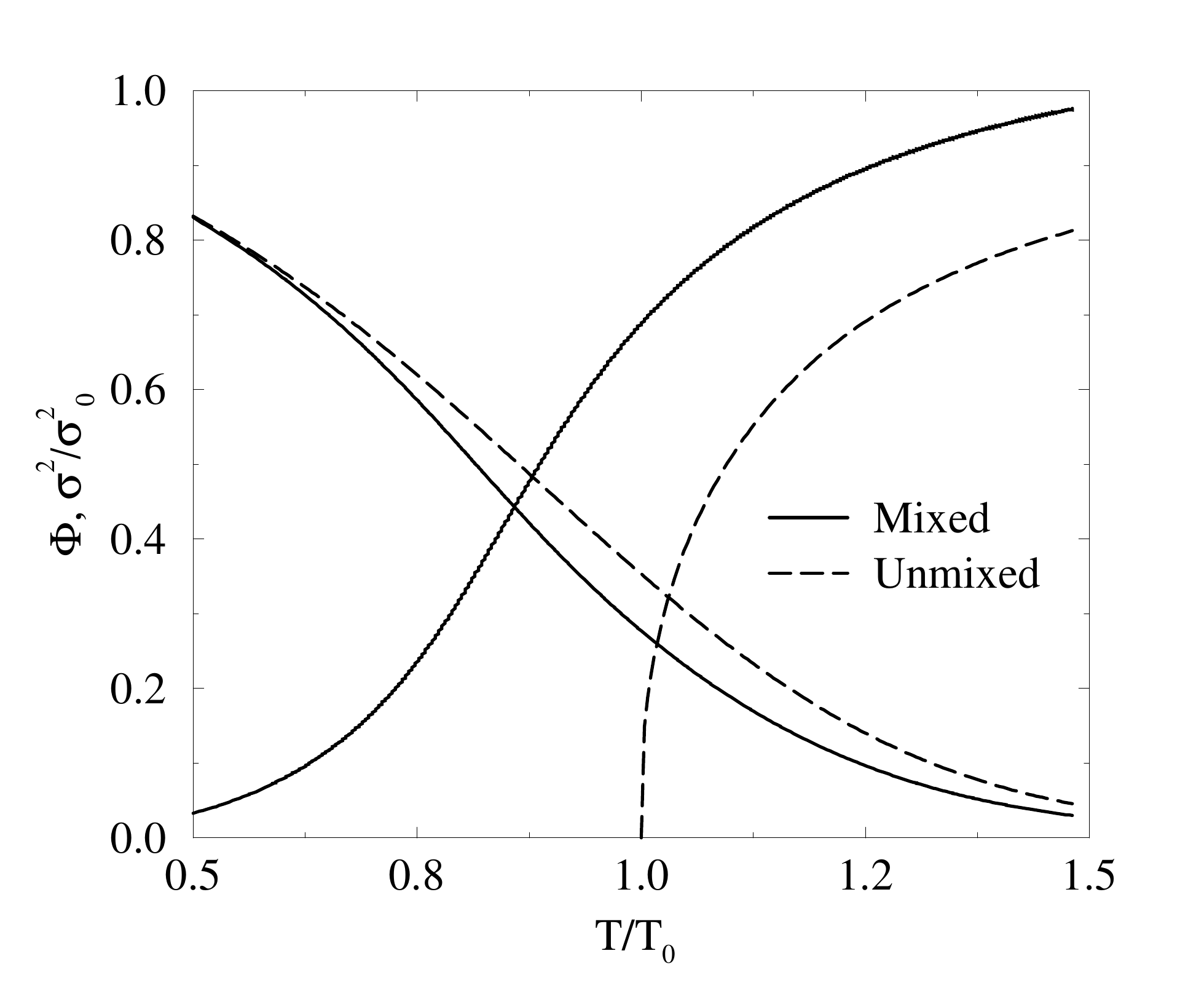}
\caption{$\Phi$ (increasing curves) and normalizaed $\sigma^2$ (decreasing ones) for $T_0=270$ MeV using $M_{\bm{k},T}$. The solid and dashed lines indicate the cases with and without the dynamical quark and $\Phi$ mixing, respectively.}       
\label{fig3}
\end{figure}

In this Section, we present the numerical results for the two different order parameters, $\sigma^{2}$ and $\Phi$ as functions of $T$, and related discussions are given as well. Hereafter, we will used the normalized value for $\sigma^{2}$, $\sigma^{2}/\sigma^{2}_{0}$ for convenience. Firstly, in Figure~\ref{fig3}, we draw them for the cases with (solid lines) and without (dashed lines) the mixing of the dynamical quark and $\Phi$, using $M_{\bm{k},T}$. Here, we employ $T_{0}=270$ MeV for the pure-glue potential in Eq.~(\ref{eq:POLPO}). In order for the unmixed case (dashed lines), in which the quarks are decoupled from the pure gluodynamics, we set $M$ infinite, equivalently $E_{{\bm k},T}\to\infty$ in Eqs.~(\ref{eq:WRP}) and (\ref{eq:WRS}). As shown in the figure, even with or without the mixing, $\sigma^{2}$ showed the crossover for the chiral phase transition. On the contrary, the phase transition pattern for $\Phi$ turns out to be different, depending on the mixing. In this sense, the crossover of $\Phi$ is caused by the mixing, as suggested by the pNJL models~\cite{Fukushima:2003fw,Ratti:2004ra,Ratti:2005jh,Ratti:2006gh,Sakai:2008py} and shown by the LQCD simulations with dynamical quarks~\cite{Bornyakov:2004ii}. An interesting behavior shown in Figure~\ref{fig3} is that the mixing effect increases then decreases for $\sigma^{2}$ as indicated in the solid and dashed lines. This tendency comes from the combination of the opposite behaviors of $\Phi$ (increasing) and $e^{-E_{\bm{k},T}/T}$ (decreasing) in $\sigma^{2}$ in Eq.~(\ref{eq:WRS}).

\begin{figure}[t]
\includegraphics[width=8cm]{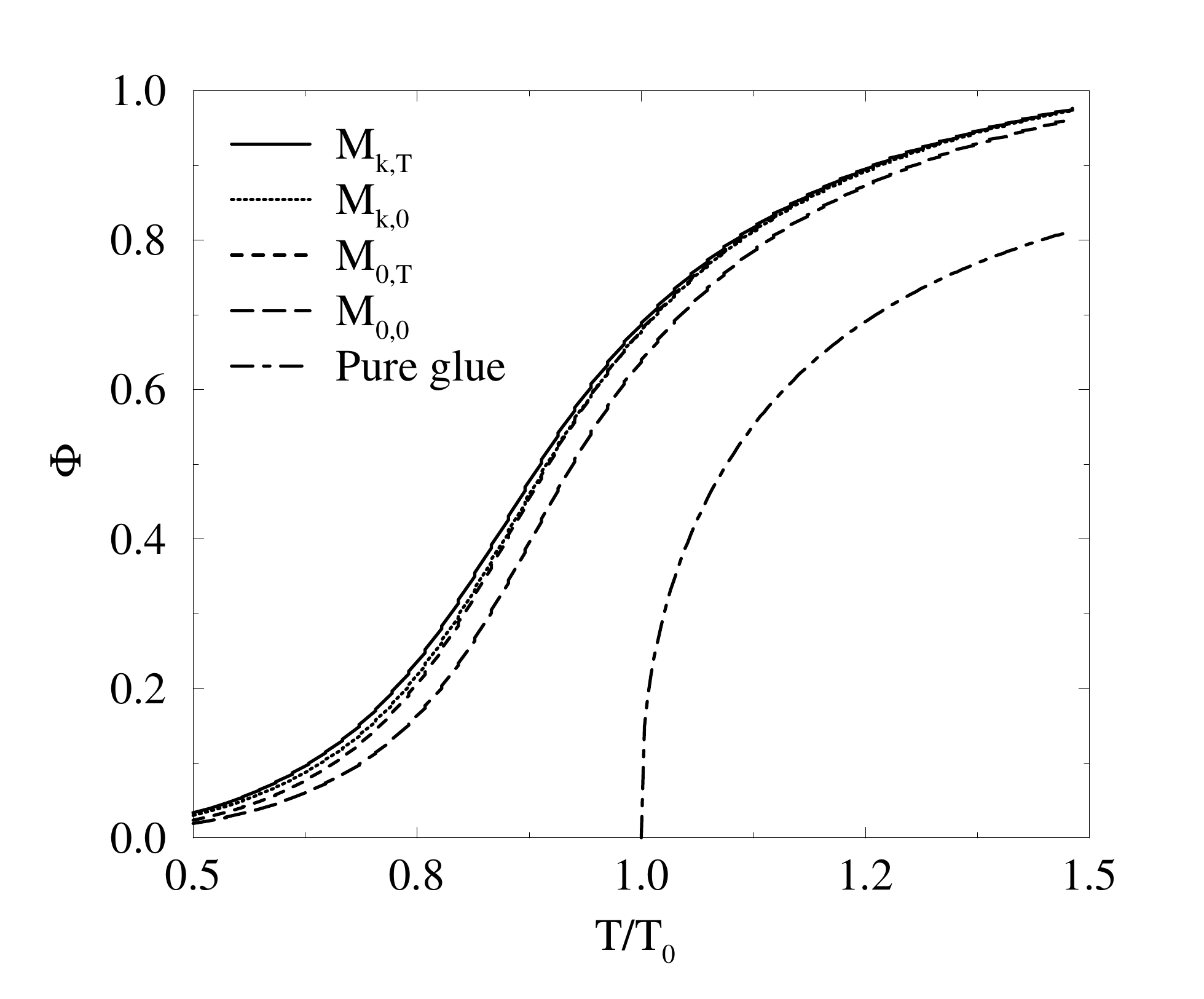}
\caption{$\Phi$ as a function of $T$ for $T_0=270$ MeV, employing $M_{\bm{k},T}$ (solid), $M_{\bm{k},0}$ (dotted), $M_{0,T}$ (dashed), and $M_{0,0}$ (long-dashed). We also draw $\Phi$ for the pure-glue case (dot-dashed).}       
\label{fig4}
\end{figure}

In Figure~\ref{fig4}, we present the numerical results for $\Phi$ as a function of $T$ for different types of $M$, as listed in Table~\ref{table2}. We again employed $T_{0}=270$ MeV. $\Phi$ for the pure-glue potential, showing the first-order deconfinment  phase transition, is also given in Figure~\ref{fig4} for comparison. From the figure, it is clearly shown that the $T$- and/or $\bm{k}$-dependent $M$ make $\Phi$ shifted to lower $T$, resulting in lowering $T_{c}$ (we will discuss the numerical values for $T_{c}$ later in detail). 
\begin{table}[b]
\begin{tabular}{c|c|c|c}
$\hspace{0.8cm}M_{\bm{k},T}\hspace{0.8cm}$
&$\hspace{0.8cm}M_{\bm{k},0}\hspace{0.8cm}$
&$\hspace{0.8cm}M_{0,T}\hspace{0.8cm}$
&$\hspace{0.8cm}M_{0,0}\hspace{0.8cm}$\\
\hline
$M_{0,T}\left[\frac{2}{2+\bm{k}^2\bar{\rho}^2}\right]^2$
&$M_{0,0}\left[\frac{2}{2+\bm{k}^2\bar{\rho}^2_0}\right]^2$
&$M_{0,0}\left[\frac{\sqrt{n}\,\bar{\rho}^2}
{\sqrt{n_{0}}\,\bar{\rho}^2_0} \right]$&$350$ MeV\\
\end{tabular}
\caption{Notations for $M$ given in Figure~\ref{fig4}.}
\label{table2}
\end{table}

In order to see the effects of $(\bm{k},T)$ dependence in $M$, we take a careful look on Eq.~(\ref{eq:WRP}). For simplicity, we approximate it considering only the leading contributions as follows:
\begin{equation}
\label{eq:WRP1}
T^3\left[b_4\,\Phi^3-b_3\,\Phi^2-b_2(T)\,\Phi\right]\approx
4N_cN_f\int\frac{d^3\bm{k}}{(2\pi)^3}
\,e^{-E_{\bm{k},T}/T}.
\end{equation}
As known from Figure~\ref{fig2} and Eq.~(\ref{eq:ENERGY}), when we take into account the $T$- and/or $\bm{k}$-dependent $M$, the strength of $M$ decreases as ${\bm k}$ and/or $T$ increases, and the same for $E_{\bm{k},T}$. As a result, the integrand in the {\it r.h.s.} of Eq.~(\ref{eq:WRP1}) tends to be larger, as ${\bm k}$ and/or $T$ increases, than that with a constant $M$, $M_{0,0}\approx350$ MeV. To make things clear, we put $T=T_{0}$ in Eq.~(\ref{eq:WRP1}) for example, then have
\begin{equation}
\label{eq:WRP2}
T^3_{0}\left[b_4\,\Phi^3-b_3\,\Phi^2\right]\approx
\underbrace{4N_cN_f\int\frac{d^3\bm{k}}{(2\pi)^3}
\,e^{-E_{\bm{k},T_{0}}/T_{0}}}_{A}
>\underbrace{4N_cN_f\int\frac{d^3\bm{k}}{(2\pi)^3}
\,e^{-E_{0,0}/T_{0}}}_{B},
\end{equation}
where we have used the notation, $E^{2}_{0,0}=M^2_{0,0}+\bm{k}^2$. In order to satisfy the relations of Eq.~(\ref{eq:WRP2}), $\Phi$ for $A$, $\Phi_{A}$ must be bigger than $\Phi_{B}$ at $T=T_{0}$, since $\Phi$ is positive definite and $b_{4}\gg b_{3}$. This observation is also true for arbitrary $T$, except for the limiting cases, $T\to0$ or $\infty$. Ats the  same time, therefore, this situation can be understood as that $\Phi$ is shifted downward almost horizontally: $T_{c}$ is lowered consequently. 

However, this lowering $T_{c}$ behavior must be understood separately for the $\bm{k}$ and $T$ dependences in $M$, since the both curves with $M_{0,T}$  and $M_{\bm{k},0}$ show it, as depicted in Figure~\ref{fig4}. 

\begin{itemize}
\item $T$ dependence: $M_{0,T}$

It tells us that the partial chiral restoration, indicated by decreasing $M\propto\bar{\rho}^{2}$ with respect to $T$, has effects on the deconfinment phase transition, although these two phase-transition mechanisms are believed to be different to each other. Here is a microscopic explanation for this: As $T$ increases, the pseudo-particle (instanton) become smaller in its size $\sim\bar{\rho}$, decreasing  QCD vacuum contribution simultaneously, resulting in that the quarks are enable to travel more freely with less interactions with the instantons, and strings attached to each quark are extended more at lower string tension, in comparison to the case without the partial chiral restoration in $M$. Hence, the condensation of the strings can happen easily at lower $T$, toward the deconfinment phase. 

\item $\bm{k}$ dependence: $M_{\bm{k},0}$

The $\bm{k}$ dependence in $M$ is originated from the delocalization of quark fields in the presence of the instanton background~\cite{Diakonov:2002fq}, not from $T$-related effects. As momentum transfer increases, quarks become lighter $\propto1/\bm{k}^{4}$, loosing the nontrivial QCD vacuum contributions. In other words, the instanton effect is weakened seemingly in the $\bm{k}$ integrals. Consequently, being similar to the $T$-dependence case, the deconfinment phase transition takes place at lower $T$.  
\end{itemize}
From these explanations, one is led to a conclusion that the QCD vacuum contributions and the partial chiral restoration play considerable roles for the deconfinment phase transition to a certain extent in the presence of the mixing. 

Now, we are in a position to calculate the critical temperatures $T_{c}$ for the crossover transitions numerically. They are determined by the chiral and Polyakov susceptibilities as in Ref.~\cite{Fukushima:2003fw}. Being almost equivalently, they can be also obtained from the maximum values of $\partial\Phi/\partial T$ and $\partial\sigma^{2}/\partial T$ for the deconfinement and  chiral phases, respectively as in Refs.~\cite{Ratti:2004ra,Ratti:2005jh,Ratti:2006gh,Rossner:2007ik}. In the present work, we employ the later method. In Table~\ref{table3}, we list them for $\Phi$ and $\sigma^2$, assigned as $T^{\mathbb{Z}}_{c}$ and $T^{\chi}_{c}$, respectively, for each type of $M$. Note that we do not show the numerical results for $T^{\chi}_{c}$ for the cases with $M_{0,T}$ and $M_{0,0}$, since they are UV divergent, proportional to $\int\bm{k}^{3}d\bm{k}$, unless a cutoff is introduced by hand. 
\begin{table}[b]
\begin{tabular}{c||c|c|c|c}
$T_0=270$ MeV&$\hspace{0.4cm}M_{\bm{k},T}\hspace{0.4cm}$
&$\hspace{0.4cm}M_{\bm{k},0}\hspace{0.4cm}$
&$\hspace{0.4cm}M_{0,T}\hspace{0.4cm}$
&$\hspace{0.4cm}M_{0,0}\hspace{0.4cm}$\\
\hline
\hline
$\Phi$&$T^{\mathbb{Z}}_{c}=227$ MeV&$T^{\mathbb{Z}}_{c}=225$ MeV
&$T^{\Phi}_c=230$ MeV&$T^{\Phi}_c=240$ MeV\\
\hline
$\sigma^2/\sigma^2_0$
&$T^{\sigma^{2}}_c=216$ MeV&$T^{\sigma^{2}}_c=265$ MeV
&$\cdots$&$\cdots$\\
\end{tabular}
\caption{$T_c$ computed from $\Phi$ and $\sigma^2/\sigma^2_0$ for $T_0=270$ MeV.}
\label{table3}
\end{table}

\begin{figure}[t]
\includegraphics[width=8cm]{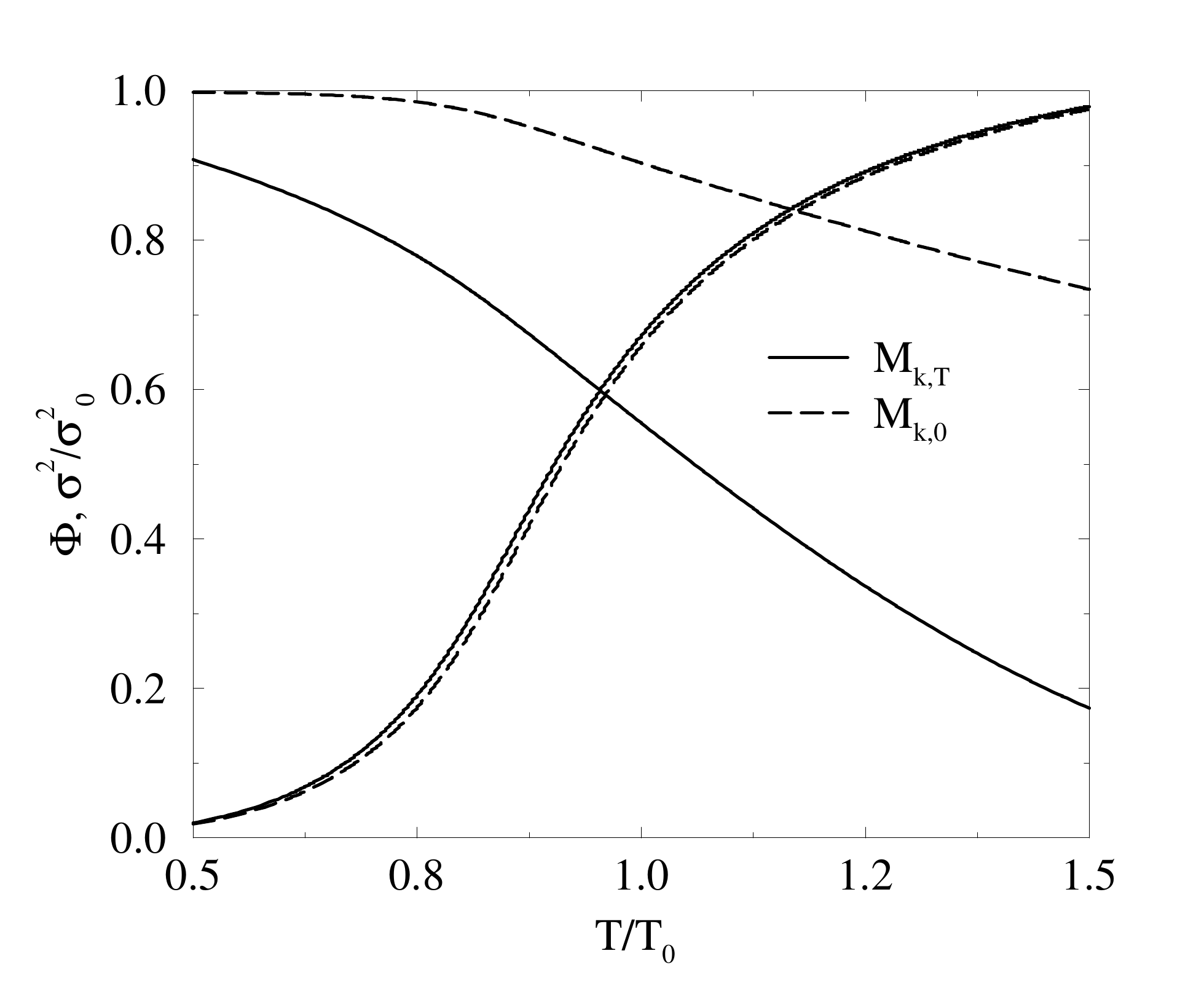}
\caption{$\Phi$ (increasing curves) and $\sigma^2/\sigma^2_0$ (decreasing ones) as functions of $T$ for $T_0=270$ MeV, employing $M_{\bm{k},T}$ (solid) and $M_{\bm{k},0}$ (long-dashed).}       
\label{fig5}
\end{figure}

As discussed previously, from the table, one can see clearly that $T_{c}$  is lowered by inclusion of the ($\bm{k},T$) dependence in $M$. It turns out that the shift of $T^{\mathbb{Z}}_{c}$ is about $10\%$, ($240\to227$) MeV for $M_{0,0}\to M_{\bm{k},T}$. Interestingly, if we take into account full ($\bm{k},T$) dependence, $T^{\mathbb{Z}}_{c}$ and $T^{\chi}_{c}$ get closer to each other as shown in Table~\ref{table3}, $(T^{\Phi},T^{\sigma^{2}})=(227,216)$ MeV. About $5\%$ discrepancy ($\sim10$ MeV) is observed between them, showing a tendency $T^{\chi}_{c}<T^{\mathbb{Z}}_{c}$. By turning off the $T$ dependence in $M$ ($M_{\bm{k},0}$), the discrepancy between $T^{\chi}_{c}$ and $T^{\mathbb{Z}}_{c}$ becomes larger up to about $15\%$, $(T^{\Phi},T^{\sigma^{2}})=(225,265)$ MeV. For interpreting this behavior, we take a look on the $T$ dependence of $\sigma^{2}$ for the cases with $M_{\bm{k},T}$ and $M_{\bm{k},0}$, and draw the numerical results in Figure~\ref{fig5}. For comparison, we also draw $\Phi$ for the two cases. Note that $\sigma^{2}$ show obvious difference for the cases with $M_{\bm{k},T}$ and $M_{\bm{k},0}$. This is the reason why the discrepancy between $T^{\mathbb{Z}}_{c}$ and  $T^{\chi}_{c}$ is so large for $M_{\bm{k},0}$. At the same time, this strong dependence on $T$ for the chiral phase transition interprets the larger shift of $T^{\chi}_{c}$, ($265\to216$) MeV. From this observation, we can conclude that the $T$ dependence in $M$ plays an important role in exploring chiral phase transition at finite $T$. 

It is worth mentioning that, from the LQCD analyses, it turned out that $T^{\chi}_{c}\approx180$ MeV for $N_{f}=2$ using the clover-improved Willson fermions~\cite{Maezawa:2007fd}. Also, using the renormalization-group (RG) improved action, it was found that $T^{\chi}_{c}\approx171$ MeV~\cite{Ali Khan:2000iz}. These values are significantly smaller than ours by $(10-20)\%$. If this is the case, one may need more strong $T$ dependence for $M$ in the present approach, since, taking a look on Eq.~(\ref{eq:WRS}), the $T$ dependence of $\sigma^{2}$ is mainly governed by the behavior of $M$. In other words, the instanton effects must decrease much faster as $T$ increases. There can be several possible scenarios to satisfy this condition:
\begin{itemize}
\item
If we consider a correct instanton distribution function, rather than the simplified one in Eq~(\ref{eq:NID2}) according to the large-$N_{c}$ limit, $T^{\chi}_{c}$ may be lowered. To test this, we have the ratio of $M$ computed with the correct one in Eq.~(\ref{eq:NID}) and $M_{0,T}$ in Eq.~(\ref{eq:momo}):
\begin{equation}
\label{eq:RATIO}
\frac{M_{\mathrm{correct}}}{M_{0,T}}=
\frac{2\mathcal{F}^{{5/2}}}{3\sqrt{\pi}\bar{\rho}^{2}}
\int d\rho\,\rho^{6}e^{-\mathcal{F}\rho^{2}}=(0.46-0.47).
\end{equation}
We, however, verify that this modification does not work for lowering $T^{\chi}_{c}$, whereas the strength of $M$ is reduced approximately by half.

\item
Additional $T$-dependent terms can be taken into account. For instance, we assumed that the Lgrangian multiplier $\lambda$ in Eq.~(\ref{eq:PF}) is independent on $T$ in deriving $M_{\bm{k},T}$ in Eq.~(\ref{eq:momo}). If it has the $T$ dependence, we can modify $M_{\bm{k},T}$ as follows: 
\begin{equation}
\label{eq: }
M_{\bm{k},T}\to\sqrt{\frac{\lambda}{\lambda_{0}}}M_{\bm{k},T}. 
\end{equation}
Quantitatively, the $T$-dependent $\lambda$ can not be determined self-consistently within the framework. From a very rough estimation, based on the assumption that $n_{0}$ in Eq.~(\ref{eq:SPE1}) converges with a brute cutoff $\Lambda\approx1/\bar{\rho}_{0}$, we can obtain a relation $\lambda_{0}\propto1/\bar{\rho}^{2}_{0}$~\cite{Diakonov:2002fq}. Using this assumption, the expression for $M_{\bm{k},T}$ in Eq.~(\ref{eq:momo}) is modified into 
\begin{equation}
\label{eq:momo2}
M_{0,T}\to
M_{0,0}\sqrt{\frac{\lambda}{\lambda_{0}}}\left[\frac{\sqrt{n}\,\bar{\rho}^2}
{\sqrt{n_{0}}\,\bar{\rho}^2_0}\right]
=M_{0,0}\left[\frac{\sqrt{n}\,\bar{\rho}}
{\sqrt{n_{0}}\,\bar{\rho}_0}\right].
\end{equation}
It turns out that this rough assumption makes things worse: $T^{\chi}_{c}$ is shifted to a larger value as expected in Eq.~(\ref{eq:momo2}). In addition, it is also possible that the reduced coefficient $\mathcal{C}$ in Eq.~(\ref{eq:IND}) and the one-loop inverse charge $\bar{\beta}$ in Eq.~(\ref{eq:para}) are functions of $T$, although it is beyond our scope in the present work.  
\item
We may not ignore the Matsubara frequency $w_{n}$ in the denominator in the {\it l.h.s.} of Eq.~(\ref{eq:MATSU}) in Appendix. In the presence of the mixing, $\Phi$ may provide effects on the instanton distribution function, becoming an exponentially decreasing function, not a gaussian one. We, however, do not perform quantitative calculations for these possibilities here and leave them for future works. 
\item
Additional $T$ dependence can be added to the instanton distribution function in Eq.~(\ref{eq:IND}) according to the fermion overlap matrix, if one considers full QCD in computing the distribution function~\cite{Ilgenfritz:1988dh}, being different from the present work based on the variational method in pure-glue QCD (the Harrington-Shepard caloron)~\cite{Diakonov:1988my}. Moreover, the instanton clustering, which was suggested as a main contribution for the chiral phase transition~\cite{Ilgenfritz:1988dh,Ilgenfritz:1994nt} and not taken into account here, may be responsible for lowering $T^{\chi}_{c}$. 
\item
We note that $T_{0}$ for the pure-glue potential can be chosen as a smaller value than $270$ MeV, which has been used throughout in the present work. As shown in Refs.~\cite{Ratti:2004ra,Ratti:2005jh,Ratti:2006gh}, by taking $T_{0}=190$ MeV, the computed values for $T^{\chi}_{c}$ became compatible with those from the LQCD simulations. Although we have not presented detailed results for lower $T_{0}$, we could obtain $T^{\chi}_{c}=194$ MeV using $M_{\bm{k},T}$ at $T_{0}=200$ MeV, showing about $10\%$ decreasing. 
\end{itemize}
As for $T^{\mathbb{Z}}_{c}$ estimated in the LQCD, using the clover-improved Wilson fermions similarly, it was determined about $210$ MeV~\cite{Bornyakov:2004ii}, which is rather compatible with ours. It was observed in the usual local pNJL model~\cite{Ratti:2004ra,Ratti:2005jh,Ratti:2006gh} that $T_{c}$ taken from the two different order parameters are almost consistent: $T^{\chi}_{c}\approx T^{\mathbb{Z}}_{c}\approx220$ MeV for $T_{0}=270$ MeV. It turned out that $T^{\mathbb{Z}}_{c}=215$ MeV in Ref.~\cite{Rossner:2007ik} with pNJL.

\begin{figure}[t]
\includegraphics[width=8cm]{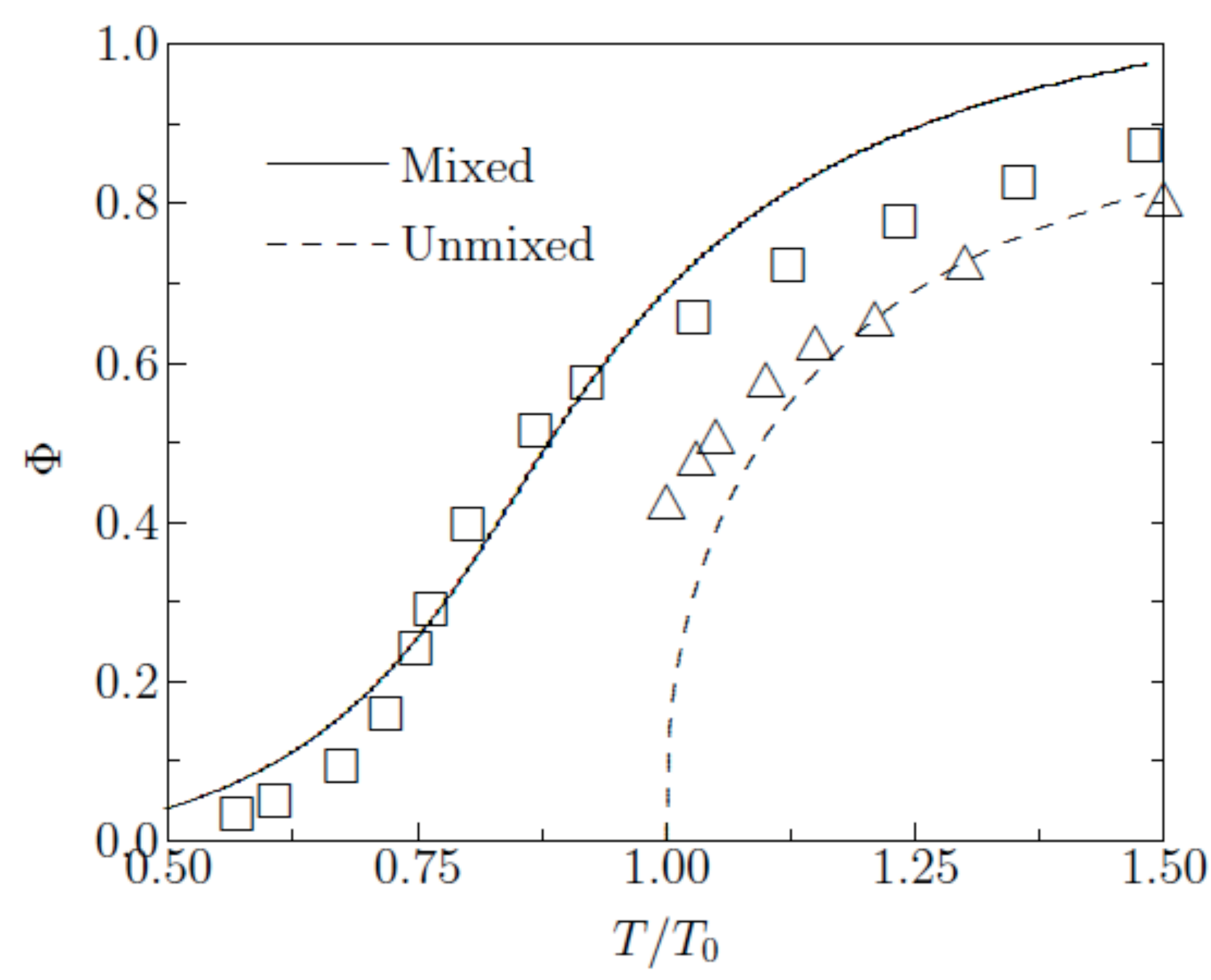}
\caption{$\Phi$ as a function of $T$ for $T_0=270$. The solid and dashed lines indicate the cases with and without the dynamical quark and $\Phi$ mixing, respectively. The notations $\Box$ and $\triangle$ indicate the full and quenched lattice data, respectively, taken from Refs.~\cite{Kaczmarek:2005ui} and \cite{Kaczmarek:2002mc}.}       
\label{fig7}
\end{figure}

Finally, we compare our results for $\Phi$ with the LQCD data from Refs.~\cite{Kaczmarek:2005ui} and \cite{Kaczmarek:2002mc} in Figure~\ref{fig7}, in which a full and quenched calculations were done for $N_{f}=2$. In their works, it was observed that $T^{\mathbb{Z}}_{c}=202$ MeV, which is about $10\%$ lower than ours, $227$ MeV for $N_{f}=2$, and $270$ MeV for $N_{f}=0$. As shown in the figure, the LQCD data, indicated by $\Box$ (full) and $\triangle$ (quenched), are in qualitative agreement with the present results, but not quantitative. Especially, our result for the mixed case deviates much from it for the region $T>T_{0}$. 

\section{Summary and conclusion}
In the present work, we have attempted to construct an effective thermodynamic potential $\Omega_\mathrm{eff}$ at finite $T$ and zero quark-chemical potential ($\mu_{\mathrm{R}}=0$) in the chiral limit.We restricted ourselves to $N_{c}=3$ and $N_{f}=2$. Motivated by the Polyakov-loop-augmented Nambu-Jona-Lasinio (pNJL) model, we wanted to incorporate two different order parameters, $\sigma^{2}$ and $\Phi$, which characterize the chiral and deconfinment phase transitions, respectively. 

In order to discuss the spontaneous breakdown of chiral symmetry at finite $T$, we employed the singular-gauge instanton solution, and the fermionic Matsubara formula to express the effective chiral action as a function of $T$. We employed the instanton-distribution function, derived from the Harrington-Shepard caloron, to obtain the instanton density and average instanton size as functions of $T$. It turned out that these two quantities decreased but finite, indicating that the instanton effect survives even beyond $T_c$~\cite{Diakonov:1988my}. The ($\bm{k},T$)-dependent $M$, $M_{\bm{k},T}$ was derived in the large-$N_c$ limit. We found that the $\bm{k}$ dependence of $M$ becomes weaker as $T$ increases. At the same time, the absolute value of $M_{\bm{k},T}$ was also reduced with respect to $T$. To include the Polyakov loop as an order parameter for the $\mathbb{Z}(N_{c})$ symmetry, we took into account imaginary quark-chemical potential $\mu_{I}\equiv A_4$, which was translated as the traced Polyakov loop $\Phi$. Combining all these ingredients, we could construct $\Omega_{\rm{eff}}$ with an additional pure-glue SU($N_{c}$) gauge potential. By minimizing $\Omega_{\rm{eff}}$ with respect to external fields such as $\sigma$ and $\Phi$, we could compute $\sigma^{2}$ and $\Phi$ as functions of $T$ numerically. From the various numerical results we have found the followings:
\begin{itemize}
\item
In the presence of the mixing of the dynamical quarks and $\Phi$, we observed that $\Phi$ is very sensitive to the mixing, showing the crosssover and first-order transitions with and without it, respectively. In contrast, $\sigma^{2}$ is insensitive to it, indicating the crossover phase transition. 
\item 
$M$ was expressed as a decreasing function of $T$ as well as $\bm{k}$. Due to this, $T^{\mathbb{Z}}_{c}$ was lowered by about $(5-10)\%$, in comparison to that with constant mass $M_{0,0}\approx350$ MeV. From this observation, we explain this lowering $T^{\mathbb{Z}}_{c}$ by that the nontrivial QCD vacuum contributions and partial chiral restoration play a significant role even in the deconfinment phase transition.  
\item 
If the ($\bm{k},T$) dependence had been fully taken into account, we found that $T^{\mathbb{Z}}_{c}=227$ MeV and $T^{\chi}_{c}=216$ MeV. The discrepancy between them became about $10$ MeV, which was rather larger than that computed in the pNJL model. We also note that the LQCD simulations presented smaller $T^{\chi}_{c}$ than ours, whereas $T^{\mathbb{Z}}_{c}$ was compatible. 
\item
Finally, we observed that $\sigma^{2}$ was depending much on $T$. Again, the partial chiral restoration turned out to be crucial to make proper results for the chiral phase transition.  
\end{itemize}

Consequently, from the present work, we could learn that the partial chiral restoration and nontrivial QCD vacuum effects must be taken into account appropriately to investigate the breaking patterns of the chiral and $\mathbb{Z}(N_{c})$ symmetries at finite $T$. As a next step, we attempt to include the finite quark-chemical potential ($\mu_{\mathrm{R}}$), giving a full description for the QCD phase diagram on the $\mu_{\mathrm{R}}$-$T$ plane. In addition to it, the finite current-quark mass, $m$ is also under consideration beyond the chiral limit. This is important, since the explicit breakdown of the flavor SU($3$) symmetry modifies the QCD phase diagram to a good extent. However, including finite $m$ into the present framework has a huddle: One needs to consider the meson-loop corrections, which make significant modification on the physical quantities such as the chiral susceptibility, in comparison to those in the chiral limit~\cite{Nam:2008bq}. Related works are underway and appear elsewhere.
\section*{Acknowledgment}
The author thanks K.~Fukushima, K.~Miura, T.~Kunihiro, and C.~W.~Kao for fruitful discussions. This work was supported by the grant of NSC96-2112-M033-003-MY3 from the National Science Council (NSC) of Taiwan. It had been also partially supported by the grant for Scientific Research (No.17070002 and No.20028005) from the Ministry of Education, Culture, Science and Technology (MEXT) of Japan. The numerical calculations were carried out on YISUN at YITP in Kyoto University. 
\section*{Appendix}
The logarithm in Eq.~(\ref{eq:TDPLP}) can be evaluated using the parametrized quark form factor, given in Eq.~(\ref{eq:FFpara}) as follows:
\begin{equation}
\label{eq:MATSU}
\ln\left[\bar{w}^2_n+\bm{k}^2+M_{0,T}^2
\left[\frac{2}{2+(\bm{k}^2+\bar{w}^2_n)\bar{\rho}^2} \right]^4 \right]
\approx\ln\left[\bar{w}^2_n+\bm{k}^2+M_{0,T}^2
\left[\frac{2}
{2+\bm{k}^2\bar{\rho}^2} \right]^4\right].
\end{equation}
We verified  that difference between the {\it r.h.s.} and {\it l.h.s.} of Eq.~(\ref{eq:MATSU}) is almost negligible. We used the following notations as
\begin{equation}
\label{eq:}
\bar{w}_n=w_n+i\mu,\,\,\,\,
M_{0,T}=M_{0,0}\left[\frac{\sqrt{n}\,\bar{\rho}^2}
{\sqrt{n_{0}}\,\bar{\rho}^2_0} \right],
\end{equation}
where $\mu$ stands for complex quark-chemical potential, $\mu\ll1$. In the present case, we choose $\mu=iA_4$. In deriving Eq.~(\ref{eq:TDPLP}) from Eq.~(\ref{eq:TDPLP2}), we also used the following Matsubara-sum indentity:
\begin{eqnarray}
\label{eq:AP1}
T\sum^{\infty}_{n=-\infty}
\ln\left[\frac{(w_n+i\mu)^2+E^2_{\bm{k},T}}{T^2}\right]
\approx E_{\bm{k},T}+T\left[\ln[1+e^{-\frac{E_{\bm{k},T}-\mu}{T}}]
+\ln[1+e^{-\frac{E_{\bm{k},T}+\mu}{T}}]\right],
\nonumber
\end{eqnarray}
where the energy of a quark is given by
\begin{equation}
\label{eq:EkT}
E_{\bm{k},T}=
\left(\bm{k}^2+M^2_{\bm{k},T}\right)^{1/2}
=\left[\bm{k}^2+M_{0,T}^2
\left[\frac{2}
{2+\bm{k}^2\bar{\rho}^2}\right]^4\right]^{1/2}.
\end{equation}


\end{document}